\documentclass[11pt]{article}
\usepackage[affil-it]{authblk} 
\usepackage{graphicx} 
\usepackage{multicol}
\usepackage{geometry}
\usepackage{blindtext}
\usepackage{parskip}
\usepackage{booktabs}
\usepackage{dirtytalk}
\usepackage{amsmath}
\usepackage{amssymb}
\usepackage{mathtools} 
\usepackage{braket}
\usepackage{xcolor}
\usepackage{subcaption}
\usepackage[backend=biber, sorting=none, maxnames=15, minnames=10]{biblatex}
\usepackage[norelsize, linesnumbered, ruled, lined, boxed, commentsnumbered]{algorithm2e}
\usepackage{overpic}
\usepackage{tikz}
\usetikzlibrary{patterns.meta}
\usetikzlibrary{decorations.pathreplacing, calligraphy}
\usepackage{dblfloatfix} 
\usetikzlibrary{positioning}
\usetikzlibrary{calc}
\usepackage[hidelinks]{hyperref}

\usepackage[dvipsnames]{xcolor}
\definecolor{dustypink}{HTML}{d48da3}
\definecolor{lightorange}{HTML}{ffcfb3}
\definecolor{butteryellow}{HTML}{fff4b3}
\definecolor{prettypink}{HTML}{ffd4d4}
\definecolor{stoneblue}{HTML}{b5d2dd}
\definecolor{darklavender}{HTML}{c197d2}

\DeclareMathOperator*{\argmax}{arg\,max}

\newenvironment{Figure}
  {\par\medskip\noindent\minipage{\linewidth}}
  {\endminipage\par\medskip}

\title{TNASS: Tensor Network Active Space Selection with the Entanglement Feature}
\author[1]{Angus Mingare\thanks{These authors contributed equally to this work.}}
\author[1,2]{Isabelle Heuz\'e\protect\footnotemark[1]}
\author[1,3]{Peter V. Coveney}

\affil[1]{Centre for Computational Science, Department of Chemistry, University College London, WC1H 0AJ, United Kingdom}
\affil[2]{Bioinformatics Institute (BII), Agency for Science, Technology and Research (A*STAR), 30 Biopolis Street, Matrix, Singapore 138671, Republic of Singapore}
\affil[3]{Advanced Research Computing Centre, University College London, WC1H 0AJ, United Kingdom}

\date{}

\begin{document}

\maketitle

\footnotetext[1]{Email: angus.mingare.22@ucl.ac.uk}
\footnotetext[2]{Email: isabelle.heuze.24@ucl.ac.uk}

\begin{abstract}
The quality of multi-scale modelling techniques in molecular electronic structure calculations, such as embedding and subspace methods, relies upon the chosen active space. The automation of active space selection is vital for ensuring the accuracy, reproducibility, and scalability in such calculations. In this work, we introduce Tensor Network Active Space Selection using the Entanglement Feature. Through the isolation of strongly correlated electrons, this method provides a scalable foundation for embedding methods in multi-scale modelling. By representing the purities of all possible orbital partitions as a Matrix Product State, our method isolates regions of strong electron correlation without requiring manual preselection of target atoms or the calculation of expensive high-order density matrices. The results demonstrate that this approach leads to lower ground state energies and more accurate dipole moments than other fully automated selection schemes such as those based solely on single-orbital entropy or the selection of spatial orbitals around the HOMO/LUMO gap.
\end{abstract}

\vspace{1mm}

\begin{multicols}{2}

\section{Introduction}

Predicting the properties of molecules from \emph{ab initio} calculations is one of the central goals of computational chemistry, with applications from drug development to materials design. In principle, the electronic structure problem is solved exactly by Full Configuration Interaction (FCI), which diagonalises the molecular Hamiltonian in the complete many-electron basis. However, the size of this basis scales combinatorially with the number of electrons and spatial orbitals, making FCI intractable for all but the smallest molecules. Hence quantum chemistry primarily deals with approximate methods that trade some accuracy for tractability. Single-reference methods build on a Hartree-Fock starting point, and aim to recover electron correlation energy for systems that are well described by a single dominant electronic configuration. Most notably coupled cluster theory with single, double, and perturbative triple excitations (CCSD(T)), is often referred to as the "gold standard" of quantum chemistry. The computational cost of CCSD(T), while still scaling as $\mathcal{O}(N^7)$ in the number of spatial orbitals, $N$, is far cheaper than FCI. However, many electronic structure problems, including transition metal complexes and homolytic bond dissociation, exhibit strong electron correlation, where a single-determinant wavefunction is insufficient because it neglects static correlation \cite{cohen2012challenges}. Such systems require multireference methods, which explicitly treat several electronic configurations as equally important, but whose cost typically increases exponentially with the number of spatial orbitals included in this treatment \cite{autocas, AVAS, AEGISS, asf}.

Active space selection is the process of identifying a specific subset of molecular orbitals to be treated explicitly as correlated, while the remaining orbitals are treated at a cheaper level of theory or frozen entirely. This allows multireference methods to remain tractable: a properly selected active space allows a relatively small set of chemically important orbitals to accurately reproduce the properties of the target system in stand-alone methods such as Complete Active Space Configuration Interaction (CASCI), extending the reach of multireference treatments to systems that would otherwise be entirely out of scope. The same principle underlies multiscale and embedding approaches, in which a chemically important active region is treated with a high level of theory while the remaining environment is described by a computationally cheaper method, extending accurate treatment to systems far larger than could be handled by any single uniform level of theory \cite{ralli2024scalable}. Active space selection is also directly relevant to quantum computing approaches to electronic structure, where the orbitals retained in the active space are mapped onto qubits; since the number of qubits available on near-term devices is severely limited, identifying the smallest active space that captures the relevant physics is essential to making quantum simulation of molecules practical \cite{bickley2025extending}.

Traditionally, active space selection relies heavily on visual inspection and chemical intuition, introducing human bias and making results difficult to validate or reproduce \cite{autocas, AVAS}. Several automated approaches have been proposed to reduce this dependence, including methods based on atomic valence spaces, orbital entropies, and atomic population analysis \cite{autocas, AVAS, AEGISS, asf}. While these approaches have significantly improved the reproducibility of active space calculations, many still depend on predefined atomic fragments or chemically motivated heuristics; these we refer to as \emph{semi-automated methods}. \emph{Fully automated} methods instead seek to apply objective, physically motivated selection criteria, reducing user bias, improving reproducibility, and lowering the technical barrier to performing multireference calculations.

Since static correlation arises from strong quantum entanglement between molecular orbitals, entanglement-based measures provide a natural indicator of which orbitals should be included in the active space. Our proposed method, Tensor Network Active Space Selection (TNASS), exploits this principle using a framework known as the Entanglement Feature (EF), which encodes the purities of all possible orbital subregions as amplitudes of a fictitious wavefunction and enables the efficient evaluation of quantum information (QI) metrics \cite{akhtar2020multiregion}. By directly encoding the entanglement structure of the electronic state, TNASS identifies strongly correlated orbital subspaces without requiring prior chemical intuition, manually selected reference orbitals, or the preselection of target atoms. Because TNASS operates directly on a tensor network representation of the electronic state, it provides a framework that may be extended to the large, strongly correlated active regions found in transition metal complexes and biochemical systems such as metalloenzyme active sites, where existing active space methods can be challenging to apply and manual selection becomes increasingly difficult.

In this work, we validate TNASS on simple first-row diatomic molecules in a non-minimal basis set, specifically we primarily present results for $BeO$ in the $6-31G^*$ basis. This system allows direct comparison against near-exact reference calculations and a controlled study of TNASS's behaviour across distinct correlation regimes within a well-characterised system, before considering extending to the larger and more chemically complex systems that motivate this work. The $6-31G^*$ basis provides a virtual space large enough for selection strategies to meaningfully differ, while keeping near-exact reference calculations affordable. Our results demonstrate that TNASS can produce lower ground state energies and more accurate dipole moments than existing fully automated selection schemes based solely on single-orbital entropy or the HOMO/LUMO gap. The remainder of this paper is organised as follows: Section 2 reviews existing active space selection methods and introduces the Entanglement Feature; Section 3 presents the TNASS methodology and its computational scaling; Section 4 validates TNASS against near-exact references and demonstrates its performance along full dissociation curves; and Section 5 discusses open questions and directions for extending TNASS to larger systems.

\section{Background}

\subsection{Active Space Selection vs. Molecular Orbital Optimisation}

It is important to distinguish between active space selection methods, which identify correlated orbitals from a pre-existing set of reference molecular orbitals, and molecular orbital optimisation methods, which modify the orbital basis itself. Since orbital optimisation can substantially improve the quality of the underlying orbitals, direct comparisons between these two categories are inherently uneven; if the initial orbitals are poorly optimised, then any selection method which can only choose from that set will inevitably produce a suboptimal active space \cite{AVAS, qicas, qio}. This motivates the integration of orbital optimisation techniques, which refine the molecular orbital basis before or after active space selection. However, in this work we focus exclusively on pure active space selection methods. The active spaces selected by TNASS may subsequently be used as the input to molecular orbital optimisation methods, such as Complete Active Space Self-Consistent Field (CASSCF), to further improve the orbital basis, but this is beyond the scope of the present work.

\subsection{Existing Active Space Selection Methods}

Existing approaches fall into three broad categories depending on the physical criteria used to identify correlated orbitals and the amount of user input required: default energy-based selection, chemically guided methods based on orbital character, and fully automated methods based on quantum information metrics.

By default, Complete Active Space Configuration Interaction (CASCI) calculations require only the number of active electrons and active spatial orbitals as input. In implementations such as PySCF, the active space is then constructed by selecting the occupied and virtual molecular orbitals closest to the highest occupied molecular orbital (HOMO) and lowest unoccupied molecular orbital (LUMO). While this provides a simple and reproducible baseline, it assumes that the most important correlated orbitals lie near the HOMO-LUMO gap which is not always the case.

Semi-automated methods incorporate chemical intuition to guide the active space selection. One widely used example is Atomic Valence Active Space (AVAS), which constructs active molecular orbitals by projecting onto a user-defined set of target atomic valence orbitals \cite{AVAS}. The molecular orbitals are rotated such that they separate into two groups: those with significant overlap with the chosen atomic orbitals and those without. The underlying assumption is that chemically important molecular orbitals are well represented by a predefined atomic valence space. Although AVAS reduces the manual effort required compared to selecting orbitals by visual inspection, it still relies on the user identifying the chemically relevant atoms or atomic orbitals beforehand. Hence, AVAS can be considered a semi-automated method rather than a fully automated active space selection procedure.

Tensor network methods offer a compact way to represent high-dimensional wavefunctions by decomposing large systems. In quantum chemistry, this allows us to efficiently calculate quantum information metrics such as orbital entropies. Fully automated active space selection methods often identify correlated orbitals directly using these quantum information metrics, removing the need for predefined atomic fragments or chemical intuition.

The most widely adopted fully automated method is perhaps AutoCAS \cite{autocas}. AutoCAS first performs a cheap (partially converged) Density Matrix Renormalization Group (DMRG) calculation to obtain an approximate ground state solution and computes the single-orbital entropy for each molecular orbital, which quantifies the degree of entanglement between an individual orbital and the remainder of the system. The spatial orbitals are then ranked according to their single-orbital entropy, and all orbitals whose entropy exceeds a fixed fraction of the maximum value, typically 10\%, are selected for the active space \cite{autocas}. For the purposes of comparison against methods parametrised by a fixed active space size, we adapt this criterion to select the $n$ spatial orbitals with the largest single-orbital entropy, rather than applying a fixed percentage threshold.

Other tensor network-based approaches combine quantum information metrics with additional chemically motivated selection criteria. Automatic Entanglement-Driven Graphical Interface for Selecting Spaces (AEGISS) \cite{AEGISS} combines the entropy-based preselection of AutoCAS with the atomic-orbital projections used in AVAS. Spatial orbitals with large single-orbital entropies are first identified, again typically using the 10\% threshold, before atomic orbital projections are applied to construct the final active space. This allows molecules to be partitioned into chemically meaningful fragments while ensuring that the selected orbitals exhibit both strong entanglement and significant overlap with user-defined atomic orbitals. AEGISS also includes a post-processing orbital optimisation step to improve the quality of the selected active space \cite{AEGISS}.

The Active Space Finder (ASF) \cite{asf} developed by HQS adopts a different quantum information metric by replacing the single-orbital entropy with the \say{two-electron cumulant}. Unlike single-orbital entropy, the cumulant directly measures electron correlation beyond the mean-field approximation and naturally extends to state-averaged calculations. 

AutoCAS serves as the primary benchmark throughout this work because it is the most widely adopted fully automated active space selection method based on quantum-information metrics. While methods such as AEGISS and ASF also employ tensor network calculations to guide active space construction, they incorporate additional selection criteria or alternative correlation measures that address related but distinct aspects of the active space selection problem. The methods discussed here are summarised in Table~\ref{tab:active_space_methods}.

\begin{table*}[t]
\centering
\caption{Comparison of active space selection methods.}
\label{tab:active_space_methods}
\resizebox{\linewidth}{!}{%
\begin{tabular}{lllll}
\toprule
\textbf{Method} &
\textbf{Automated} &
\textbf{User Input} &
\textbf{Quantum Information Metric} &
\textbf{Orbital Optimisation} \\
\midrule
Default CASCI & Yes & Number active orbitals & None & No \\
AVAS & Semi & Target atomic orbitals & None & Yes \\
AutoCAS & Yes & None (optionally number active orbitals) & Single-orbital entropy & No \\
AEGISS & Semi & Atomic fragments / target AOs & Single-orbital entropy & Yes \\
ASF & Yes & None & Two-electron cumulant & No \\
TNASS & Yes & None (optionally number active orbitals) & Multi-orbital entropy & No \\
\bottomrule
\end{tabular}
}
\end{table*}

\subsection{The Entanglement Feature}

The Entanglement Feature (EF) characterises many-body systems by encoding bipartite entanglement as amplitudes within a fictitious wavefunction \cite{ef}. This is useful for active space selection as entanglement measures provide a basis for identifying which orbitals are important for multi-configurational descriptions \cite{autocas}.

Consider an $L$-site system, where $L$ is the number of molecular spin-orbitals, partitioned into two complementary subsystems, $A$ and $B$ as shown in Figure~\ref{fig:blob}. In second quantisation notation a bitstring, $\ket{b} \in \{\ket{0}, \ket{1}\}^L$, can be used to label these partitions where a value of \say{1} indicates that a site is in group $A$ and \say{0} indicates that a site is in the complement of $A$ (that is, group $B$).

\begin{Figure}
    \centering    \includegraphics[width=0.6\linewidth]{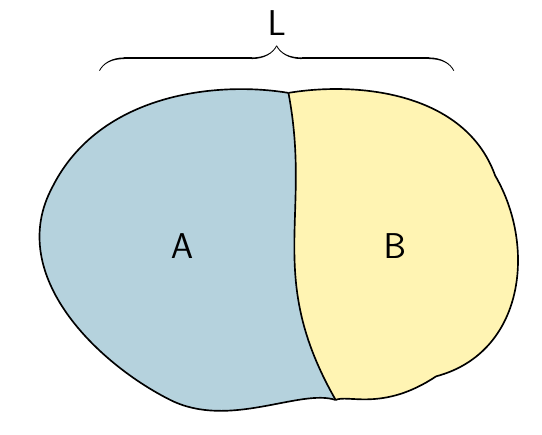}
    \captionof{figure}{A representation of an $L$-site quantum system partitioned into two complementary, bipartite subregions, $A$ and $B$.}
    \label{fig:blob}
\end{Figure}

The (unnormalised) EF is then defined to be

\begin{equation}
    \ket{EF} = \sum_{b\in{\{0,1\}}^L} e^{-S_2(b)} \ket{b},
\end{equation}

where $S_2(b)$ is the Réyni-2 entropy of the bipartition defined by $\ket{b}$,

\begin{equation}
    S_2(b) = -\log\mathrm{Tr}(\rho_A^2).
\end{equation}

The Réyni-2 entropy is used here so that the amplitudes in the EF are simply subsystem purities which are easily calculated with tensor network contractions. We provide an overview of the tensor network construction described by Kolisnyk et al. in Section 3.1. 

In the context of quantum chemistry calculations, the system of interest is described by $L$ molecular spin-orbitals which must be partitioned into an active space and an environment. Because the Réyni-2 entropy can be used as a proxy for entanglement, the EF can serve as an oracle to rank any subset of orbitals based on their degree of entanglement with the rest of the system \cite{ef, akhtar2020multiregion, AEGISS}. In this way, we develop automated active space selection schemes which go beyond the single-orbital entanglement measure of AutoCAS and account for genuinely multi-orbital entanglement.

\section{Methods}

\subsection{Building the Entanglement Feature}

Tensor networks provide a natural framework for automated active space selection methods as they enable efficient evaluations of quantum information metrics. Often, these methods begin by performing a cheap DMRG calculation to obtain an approximate ground state solution, $\ket{\psi}$, represented as a Matrix Product State (MPS) as shown in Figure~\ref{fig:mps}. 

\begin{Figure}
    \centering    \includegraphics[width=0.3\linewidth]{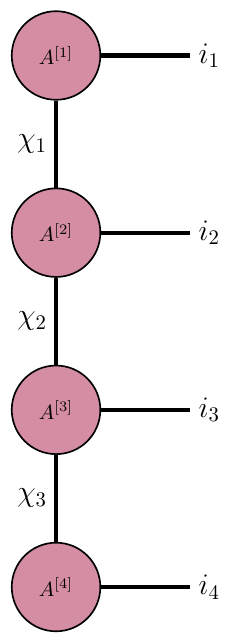}
    \captionof{figure}{An MPS is a one-dimension chain of tensors, $A^{[i]}$, representing a quantum state. The physical dimension of site $k$ is $i_k$.}
    \label{fig:mps}
\end{Figure}

An MPS is a representation of a quantum state by a one-dimensional chain of tensors that when contracted together reproduce the original state. Each external index represents a physical degree of freedom (in our case, a single spin-orbital). The dimension of the internal indices, $\chi$ (known as the \say{bond dimension}), controls the accuracy and tractability of calculations involving tensor networks. 

Given the MPS representation of $\ket{\psi}$ we can construct an MPS representation of the EF, $\ket{EF}$, using only local operations. This is possible since for a partition of the system into $\{i\}$ and its complement we have,

\begin{equation}
    S_2(\{i\}) = \braket{\psi |\braket{\psi|\mathbb{I}- \text{SWAP}_{\{i\}}|\psi}|\psi},
    \label{eq:swap_entropy}
\end{equation}

where the $\text{SWAP}$ operator acts on sites $\{i\}$ across two copies of the state $\ket{\psi}$. Equation~\ref{eq:swap_entropy} is true whenever each site of the MPS corresponds to a spin-orbital of the system which holds for DMRG run on the second-quantised fermionic Hamiltonian or, in our case, a Jordan-Wigner encoded qubit Hamiltonian. To construct the EF, we proceed in the following way. 

Given each MPS tensor $A^{[i]}$ (Figure~\ref{fig:tensor}), we first form a doubled tensor by taking the Kronecker product of two copies (Figure~\ref{fig:copied_tensor}) of the local tensor, 

\begin{equation}
    D^{[i]} = A^{[i]} \otimes A^{[i]},
\end{equation}

such that the virtual bond dimensions are squared. For bulk tensors, the resulting object has the form

\begin{equation}
    D^{[i]}_{(u_1u_2)(d_1d_2)p_1 p_2} = A^{[i]}_{u_1d_1p_1} A^{[i]}_{u_2d_2p_2},
\end{equation}

as shown in Figure~\ref{fig:double_ds}.

\begin{Figure}
    \centering    \includegraphics[width=0.5\linewidth]{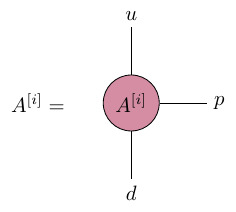}
    \captionof{figure}{One site of the MPS, $A^{[i]}$ with indices $u$ (up), $d$ (down), and $p$ (physical).}
    \label{fig:tensor}
\end{Figure}

\begin{Figure}
    \centering    \includegraphics[]{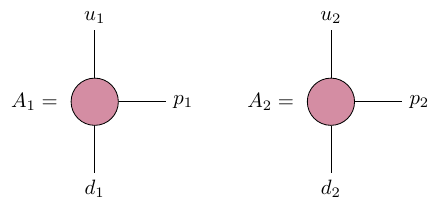}
    \captionof{figure}{Visual depiction of two identical copies ($A_1$ and $A_2$) of the local tensor $A^{[i]}$.}
    \label{fig:copied_tensor}
\end{Figure}

\begin{Figure}
    \centering    \includegraphics[width=0.9\linewidth]{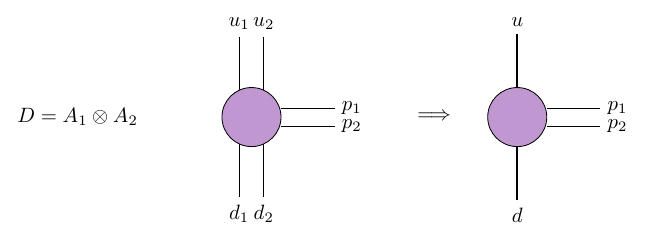}
    \captionof{figure}{Construction of the doubled tensor $D = A_1 \otimes A_2$ via the Kronecker product of two local tensor copies, resulting in squared virtual bond dimensions.}
    \label{fig:double_ds}
\end{Figure}

Two distinct tensor contractions are constructed formed of $D^{[i]}$ and its conjugate, $D^{[i]\dagger}$, to make the identity tensor, $TN_{\text{id}}$ (Figure~\ref{fig:TN_id}) and the swap tensor, $TN_{\text{swap}}$ (Figure~\ref{fig:TN_swap}).

\begin{Figure}
    \centering    \includegraphics[width=0.45\linewidth]{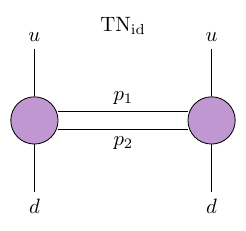}
    \captionof{figure}{The identity tensor network contraction ($T_{\text{id}}$), where physical indices of the doubled state are mapped and contracted directly without permutation.}
    \label{fig:TN_id}
\end{Figure}

\begin{Figure}
    \centering    \includegraphics[width=0.45\linewidth]{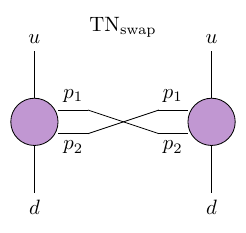}
    \captionof{figure}{The swap tensor network contraction ($T_{\text{swap}}$), where physical indices are permuted prior to contraction to compute the swap operator expectation value.}
    \label{fig:TN_swap}
\end{Figure}

In the identity contraction, physical indices are contracted directly. In the swap contraction, physical indices are switched prior to contraction. After contraction over physical indices, the two up virtual indices are combined and the two down virtual indices are combined. This results in two rank-2 tensors, $T_{\text{id}}$ and $T_{\text{swap}}$ (Figure~\ref{fig:two_Ts}), each with bond dimension $\chi^4$, where $\chi$ is the original MPS bond dimension.

\begin{Figure}
    \centering    \includegraphics[width=0.5\linewidth]{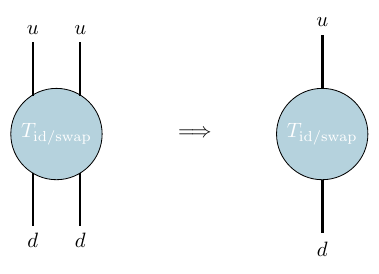}
    \captionof{figure}{After contracting $TN_{\text{id}}$ ($TN_{\text{swap}}$) we obtain $T_{\text{id}}$ ($T_{\text{swap}}$).}
    \label{fig:two_Ts}
\end{Figure}

These two objects, $T_{\text{id}}$ and $T_{\text{swap}}$, are the components of the rank-3 local tensor of the EF, $T^{[i]}$,

\begin{equation}
    T^{[i]}_{ud0} = T_{\text{id}},
\end{equation}
\begin{equation}
    T^{[i]}_{ud1} = T_{\text{swap}},
\end{equation}

defining the MPS representation of the entanglement feature (Figure~\ref{fig:final_T}).

\begin{Figure}
    \centering    \includegraphics[width=0.5\linewidth]{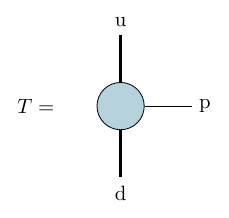}
    \captionof{figure}{The final tensor, $T$, with bond dimension $\chi^4$, which tracks the internal states across the network to build the EF-MPS representation.}
    \label{fig:final_T}
\end{Figure}

The entanglement feature MPS (EF-MPS) is contracted with binary bitstrings representing subsystem partitions. For a given partition, $\ket{b}$,

\begin{equation}
    \ket{b} = \ket{b_1b_2\cdots b_L} \quad ; \quad b_i \in \{0,1\},
\end{equation}

the EF contraction evaluates,

\begin{equation}
    \braket{b|EF} = e^{-S_2(b)}.
\end{equation}

Therefore, we obtain the Rényi-2 entropy associated with the chosen partition with a simple tensor network contraction,

\begin{equation}
    S_2(b) = -\log\braket{b|EF}.
\end{equation}

\subsection{Bond Dimensions}

The approximate ground state MPS for the target system should be obtained from a cheap (low bond dimension) DMRG calculation so that it does not dominate the runtime of the calculation. However, to be useful for active space selection (both AutoCAS and TNASS), the MPS should capture enough of the system's entanglement to accurately select correlated orbitals. In particular, a minimum requirement should be that the MPS can recover the qualitative features of the single-orbital entropy distribution. 

TNASS faces a second, related challenge. We use multi-orbital entanglement obtained from the EF-MPS to select an active space. As noted in Section 3.1, the bond dimension of the EF-MPS is $\chi^4$ where $\chi$ is the bond dimension of the original MPS. Thus, only for very small bond dimensions is TNASS feasible as demonstrated in Table~\ref{tab:ef_feasibility}.

\begin{table*}[ht]
\centering
\caption{Estimated memory cost of direct EF construction versus DMRG bond dimension $D$ for $BeO / 6-31G^*$ (56 spin-orbital sites). Memory estimates are a lower bound based on $\mathcal{O}(D^2)$ storage per site with $D = \chi^4$, assuming complex128 precision and an 8.0\,GB budget.}
\label{tab:ef_feasibility}
\begin{tabular}{llll}
\toprule
$\chi$ & EF bond dim ($\chi^4$) & Est.\ memory (GB) & Verdict \\
\midrule
2  & 16              & $<0.001$           & Safe   \\
4  & 256             & 0.055              & Safe   \\
6  & 1{,}296         & 1.402              & Safe   \\
8  & 4{,}096         & 14.000             & Unsafe \\
16 & 65{,}536        & 3{,}584            & Unsafe \\
64 & 16{,}777{,}216  & $2.3 \times 10^8$  & Unsafe \\
\bottomrule
\end{tabular}
\end{table*}

\begin{figure*}
    \centering    \includegraphics[width=\textwidth]{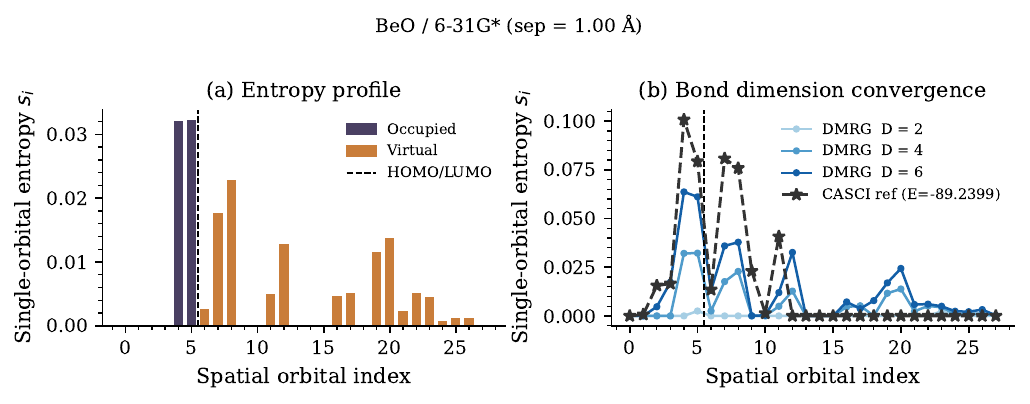}
    \captionof{figure}{The single-orbital entropy distribution of the approximate ground state MPS at bond dimension $\chi=2,4,6$ compared to a reference CASCI(12e, 12o) calculation for the $BeO/6-31G^*$ system at separation distance $0.8\AA$.}
    \label{fig:bd_convergence}
\end{figure*}

The utility of TNASS comes from the observation that relatively low bond dimensions are sufficient to recover the features of the system's entanglement. As demonstrated in Figure~\ref{fig:bd_convergence}, even with bond dimensions of $\chi = 4$ or $\chi=6$, the resulting single-orbital entropy curves exhibit very close qualitative behaviour to our benchmark (obtained with a CASCI(12e, 12o) calculation), with peaks and troughs at approximately the same locations, despite differences in magnitude.

Note that in Figure~\ref{fig:bd_convergence}, which depicts the single-orbital entropies for $BeO/6-31G^*$, we can only compare spatial orbitals with indices $2$ to $13$ since only these 12 orbitals were included in the reference CASCI calculation.

For larger systems, it is possible that such a low bond dimension MPS does not capture enough of the system's entanglement to produce good active spaces. To address this, a more expensive (though still cheap relative to the rest of the workflow) DMRG may be run at a higher bond dimension to capture more entanglement information before truncating the resulting MPS to a low bond dimension for EF construction. Alternatively, Kolisnyk et al. \cite{ef} construct compressed EFs using Tensor Cross Interpolation (TCI) and achieve significant scaling reductions across a range of test cases compared to the direct construction. However, we do not consider this in the present work. We find that the direct construction with $\chi=4$ is sufficient to produce appropriate active spaces for our test systems. 

\subsection{TNASS}

Finally, we present the active space selection strategies using the EF construction. The computational scaling of these methods is deferred to Section 3.4.

Given a target system, Hartree–Fock (HF) molecular orbitals are first obtained through restricted HF calculations. From these we generate the second-quantised molecular Hamiltonian. A DMRG calculation is run with a low bond dimension to obtain an approximate ground state MPS from which we construct the EF as in Section 3.1.

To identify an optimal active space of size $n$, we use the EF representation of the MPS as an entropy oracle. The aim is to select a subset of spatial orbitals $A$ (where $|A| = n$) that maximizes the Rényi-2 entropy, $S_2(A)$, across the bipartition between $A$ and its complement, $\bar{A}$. 

The most straightforward approach is an exact combinatorial search. In this \say{brute force} approach, we evaluate every possible subset $A \subset \{1, \dots, N\}$ of size $n$, where $N$ is the total number of spatial orbitals. The Rényi-2 entropy $S_2(A)$ is evaluated for each combination, and the subset yielding the global maximum entropy is selected. While this method is guaranteed to find the global optimum, it scales combinatorially as $\binom{N}{n}$, which makes it intractable for large numbers of orbitals or active spaces.

One alternative to this exact search is our \textbf{greedy EF method}. This begins with an empty active space $A = \emptyset$. At each iteration, the remaining orbital $i \notin A$ that gives the maximal increase in the subset entropy is added to the active space:

\begin{equation}
A_{t+1} = A_t \cup \{\argmax_{i \notin A_t} S_2(A_t \cup \{i\}) \}.
\end{equation}

This process repeats iteratively until $|A| = n$. While more computationally efficient than the brute force method, choices cannot be amended or backtracked once selected and it can miss orbitals with high single entropy which could potentially be important for the system.

The \textbf{block greedy EF method} mixes the single-orbital ranking and the greedy approach by introducing a new parameter $k$ which defines the size of each block. The algorithm begins by selecting the spatial orbital with the highest single-orbital entropy to initialise the first block. It then continues with a standard greedy selection to build up a local subset. However, once the number of selected orbitals reaches $k$, the method resets. The algorithm then selects the unchosen orbital with the highest single-orbital entropy to initialise the next block, then reapplies the greedy method. This process repeats until a total of $n$ spatial orbitals have been chosen. This method interpolates between the single-orbital entropy ranked active space ($k=1$) and the greedy EF active space ($k=n$). 

Finally, the \textbf{best $k$ greedy EF method} returns active space selections for all $k=1,...,n$. We then select the best active space to be the one which produces the lowest energy in a CASCI calculation. While we are selecting the best $k$ manually with respect to energy, the resulting active space - once determined by the energy - can be reused for further calculations such as the dipole moment.

As stated above these methods require an input parameter, namely the desired number of active spatial orbitals, $n$. However, these approaches can be implemented with a default cutoff parameter, $\epsilon$. If the addition of the next selected orbital does not increase the multi-orbital entanglement by at least $\epsilon$ (as measured by the EF) then the algorithm terminates. For example, $\epsilon$ may be a function of total single-orbital entropy, similar to the fully automated variants of AutoCAS. In all our comparisons, however, we fix $n$ to a specific value for both TNASS and AutoCAS. 

We note here that we additionally attempted to use the EF as a scoring function within a simulated annealing approach using a Metropolis-Hastings random walk over the state space consisting of all valid orbital subsets of fixed size $|A| = n$. However, we found that even with a well-calibrated annealing schedule and warm starting the initial point this method was largely unreliable. As such, we do not report its performance in this work. 

\subsection{Computational Scaling}

All the TNASS methods share a pre-processing step: the single-orbital entropies are computed directly from the DMRG MPS using single-site reduced density matrix contractions at a cost of $\mathcal{O}(N\chi^2)$ where $N$ is the number of spatial orbitals and $\chi$ is the original MPS bond dimension. This step is required by every method and only occurs once. Single-orbital entropy selection (AutoCAS) requires nothing beyond this; the active space is determined by simply ranking the single-orbital entropies with no EF contractions. It can be understood as the $k=1$ limit of the block greedy family, in which each block contains exactly one orbital chosen by the single-orbital entropy ranking and no greedy extension is performed, giving zero EF oracle calls. 

For $k>1$, the block greedy method performs additional evaluations of the multi-orbital Réyni-2 entropy $S_2(b)$ via contraction of the EF-MPS. Each such \say{oracle call}, costs $\mathcal{O}(L\chi^4)$ where $L$ is the total number of sites (number of spin orbitals). The total number of oracle calls for block greedy EF with restart interval $k$, targeting an active space of size $n$, is

\begin{align}
    &C_{\text{block}}(k,n;N) = \sum_{p=1}^{n-1} (N-p) - \sum_{q=1}^{\lceil \frac{n}{k} \rceil -1} (N-qk) \\ 
    &= (n-1)\left( N - \frac{n}{2} \right) - \left({\huge\lceil} \frac{n}{k} {\huge\rceil} - 1 \right) \left( N - \frac{\lceil \frac{n}{k}\rceil k}{2} \right).
\end{align}

This can be simplified when $k$ exactly divides $n$, 

\begin{equation}
    C_{\text{block}}(k,n;N) = \left( \frac{k-1}{k} \right) \left( Nn - \frac{n^2}{2} \right),
\end{equation}

which vanishes at $k=1$ and recovers the greedy cost $(n-1)(N - n/2)$ at $k=n$. The overhead relative to single-orbital entropy selection is therefore $C_{\text{block}}(k,n;N)$ oracle calls, each at cost $\mathcal{O}(L\chi^4)$ which gives a total additional cost $\mathcal{O}(\frac{k-1}{k} nNL\chi^4)$ to leading order. For any $k<n$ this is strictly less than the full greedy cost and quantifies the tradeoff required to exploit multi-orbital entanglement structure. 

The best $k$ method has leading order complexity $C_{\text{best $k$}} \approx \mathcal{O}(n^2N)$ where the additional overhead factor of $\mathcal{O}(n)$ arises from evaluating all $k$ values simultaneously. We reduce the prefactor of the total cost by first computing the full greedy path once as a shared resource before evaluating the remaining blocks for all $k=2,...,n-1$ independently.

A demonstration of these scaling behaviours for the $CO / 6-31G^*$ system is shown in Figure~\ref{fig:runtime_comparison}.

\begin{Figure}
    \centering    \includegraphics[width=\linewidth]{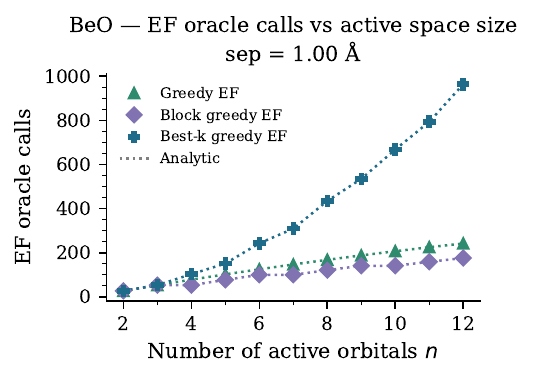}
    \captionof{figure}{Number of EF oracle calls as a function of active space size $n$ for each selection method. Coloured markers show measured call counts and black crosses with dotted lines show the analytic predictions.}
    \label{fig:runtime_comparison}
\end{Figure}

\section{Results}

\subsection{Implementation Details}

Hartree–Fock and CASCI calculations were performed using PySCF \cite{pyscf}. All tensor network calculations, including DMRG and the construction and contraction of the Entanglement Feature, were carried out using TN4QA, an in-house package available on PyPI. The DMRG functionality in TN4QA offloads to Block2 \cite{block2}. The DMRG calculations to obtain the approximate ground state MPS were run for 50 sweeps using the default Block2 convergence threshold of $10^{-8}$. To improve convergence and ensure smooth behaviour along each dissociation curve, we adopted a standard parameter transfer strategy in which the converged MPS at separation distance $r_i$ was used to initialise the DMRG calculation at the neighbouring geometry $r_{i+1}$, rather than initialising each geometry independently from a fresh Hartree–Fock reference state. Orbital selection was performed using MPS obtained at bond dimension $\chi=4$, following the observation in Section 3.2 that this is sufficient to capture the qualitative entanglement structure required for active space selection. We acquire reference values using CCSD(T) calculations performed with PySCF. All calculations were performed locally on a personal laptop.

We selected $BeO/6-31G*$ as our primary demonstration system for several reasons. First, its homolytic dissociation is a well-characterised example of strong static correlation in the quantum chemistry literature, arising from the near-degeneracy of orbitals with $\sigma$ and $\pi$ character as the bond stretches, making it a natural testbed for active space selection methods. Second, at $6-31G*$ the system has 56 spin-orbital sites, large enough that the full active space is intractable for brute-force combinatorial search yet small enough that a CCSD(T) reference remains computationally affordable, allowing direct validation of the selected active spaces against a near-exact benchmark. Third, the asymmetry between the beryllium and oxygen centres produces qualitatively different entanglement structure across the dissociation curve, as shown in Figure~\ref{fig:lilguy}, giving compressed, intermediate, and stretched geometries with distinct correlation character within a single system and allowing us to test our methods across regimes without needing to vary the system.

\subsection{Methods Validation}

Based on the given active space size and number of correlated electrons, the CASCI solver by default takes the highest occupied and lowest unoccupied orbitals to form the active space \cite{pyscf}. While selecting the highest occupied and lowest unoccupied molecular orbitals around the HOMO/LUMO gap works well for simple molecules, we expect it to be suboptimal for more complex systems which are characterised by strong static correlation, where important orbitals lie outside this narrow energy window.

To investigate this we start by considering the dissociation of diatomic systems and present the results for $BeO/6-31G^*$. We first plot the single-orbital entropy as a function of orbital index and separation distance (Figure~\ref{fig:lilguy} to identify points along the dissociation where electronic behaviour is diffuse, that is, not concentrated around the HOMO-LUMO gap.

\begin{Figure}
    \centering    
    \includegraphics[width=0.95\linewidth]{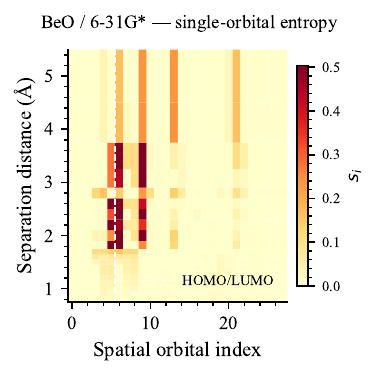}
    \captionof{figure}{The single orbital entropy, $s_i$, as a function of orbital index and separation distance for $BeO/6-31G^*$.}
    \label{fig:lilguy}
\end{Figure}

From Figure~\ref{fig:lilguy} we observe that the orbitals around the HOMO-LUMO gap dominate for intermediate separation distances ($1.8\AA$ to $3.8\AA$), whereas for smaller or larger separation distances orbitals away from the HOMO-LUMO gap may be more relevant to the electronic behaviour. This is consistent with the fact that electron correlation effects are smallest at near-equilibrium geometries. Therefore, we suspect that HOMO-LUMO-based active space selections are likely to be sufficient for these intermediate geometries. Conversely, for compressed or stretched geometries, our algorithms may work better to identify the critical active orbitals required to describe the resulting correlation. Accurate descriptions of these regimes are important for the characterisation of the transition states of chemical reactions and other dynamic processes. 

The results that follow compare the active spaces found by TNASS methods (namely greedy, block greedy with $k=3$, and best $k$ greedy) against those found with the automatic orbital selection of PySCF's CASCI solver, labelled \say{auto}, as well as with a selection criterion that uses only the $n$ orbitals with the highest single-orbital entropy, labelled \say{single orbital entropy}. While we have not explicitly implemented AutoCAS, the orbitals are ranked by the single orbital entropy and so will be in the same order as ones calculated using the AutoCAS method \cite{autocas}.

\begin{figure*}
    \centering    
    \includegraphics[width=\linewidth]{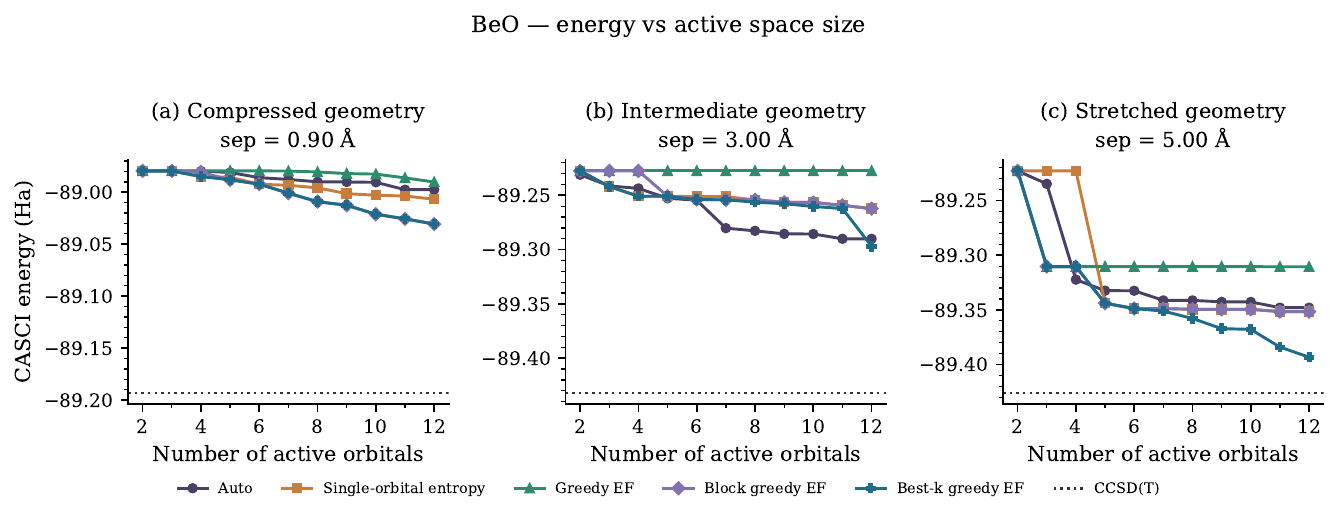}
    \caption{(a) Compressed geometry, (b) intermediate geometry, (c) stretched geometry for $BeO/6-31G^*$. All three plots show CASCI energy as a function of active space size for automatically selected orbitals (navy circles), single-orbital entropy ranked orbitals (orange squares), greedy EF selected orbitals (green triangles), block greedy EF selected orbitals (purple diamonds), and best-$k$ greedy EF selected orbitals (blue pluses). The dotted line is a reference CCSD(T) calculation.}
    \label{fig:energy_vs_active_size}
\end{figure*}

For a compressed, intermediate, and stretched geometry we perform a sweep and calculate the CASCI energy for each chosen active space as a function of the active space size, $n$. We also include a reference CCSD(T) calculation. These results are shown in Figure~\ref{fig:energy_vs_active_size}. 

Across the three representative geometries of the $BeO/6-31G^*$ dissociation curve, the entanglement feature methods track the underlying entanglement structure revealed by the heatmap. They outperform the default HOMO/LUMO selection at both the compressed geometry ($0.9\AA$) and the stretched geometry ($5.0\AA$), where correlation is diffuse across many orbitals, while being matched or exceeded by the default selection at the intermediate geometry ($3.0\AA$), where correlation is concentrated near the HOMO/LUMO gap and the assumption underlying the default method happens to hold. Among the entanglement feature methods, greedy selection is consistently the weakest performer, plateauing early and failing to approach the DMRG ground truth even at the largest active space sizes tested, a pattern consistent with the algorithm becoming trapped in a local optimum that its non-backtracking choices cannot escape. Block greedy and best-$k$ both closely track or improve upon single-orbital entropy selection across all three geometries, with best-$k$ proving the strongest and most consistent of the two: it matches block greedy at the compressed geometry, edges ahead of both single orbital entropy and block greedy at the intermediate geometry (particularly at large active space sizes, where it overtakes the otherwise dominant default selection), and remains the best performing method throughout the stretched geometry. 

These gains come at increased computational cost relative to single-orbital entropy ranking, since the entanglement feature methods require explicit evaluation of multi-orbital correlations that the single-orbital method ignores entirely. For a small diatomic like $BeO$ this overhead is modest, but the underlying failure mode of single-orbital entropy, that is its blindness to correlations that are not captured by any individual orbital's marginal entropy, should only become more pronounced in larger and more strongly correlated systems, where the relevant active space cannot be identified by proximity to the HOMO/LUMO gap alone and cheap marginal metrics are correspondingly more likely to miss the orbitals that jointly carry the important entanglement.

\subsection{Dissociation Curves}

\begin{figure*}[t]
    \centering

    \begin{subfigure}{0.48\textwidth}
        \centering
        \includegraphics[width=\linewidth]{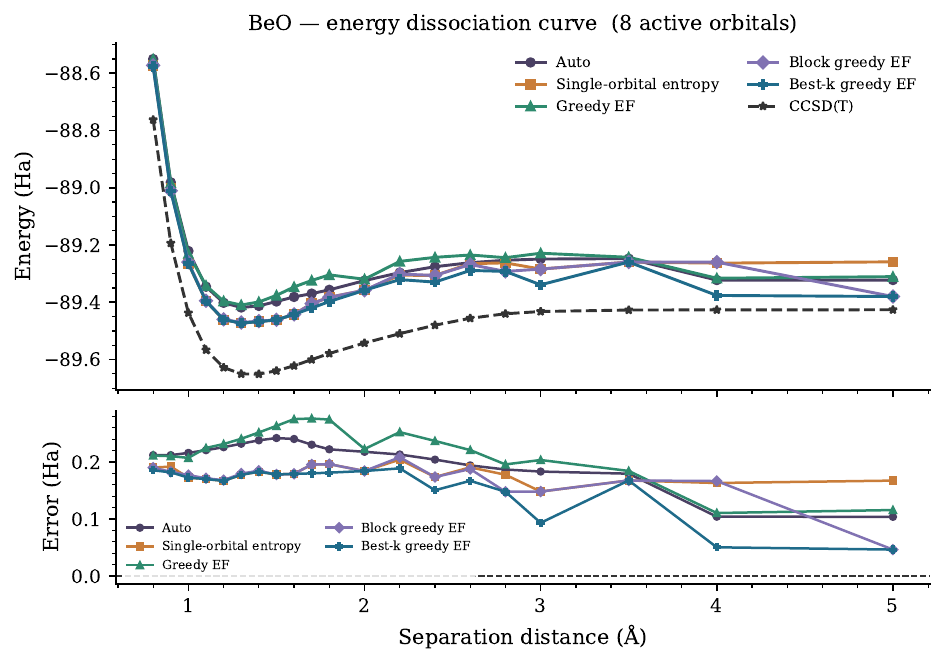}
        \caption{}
        \label{fig:energy_curve1}
    \end{subfigure}
    \hfill
    \begin{subfigure}{0.48\textwidth}
        \centering
        \includegraphics[width=\linewidth]{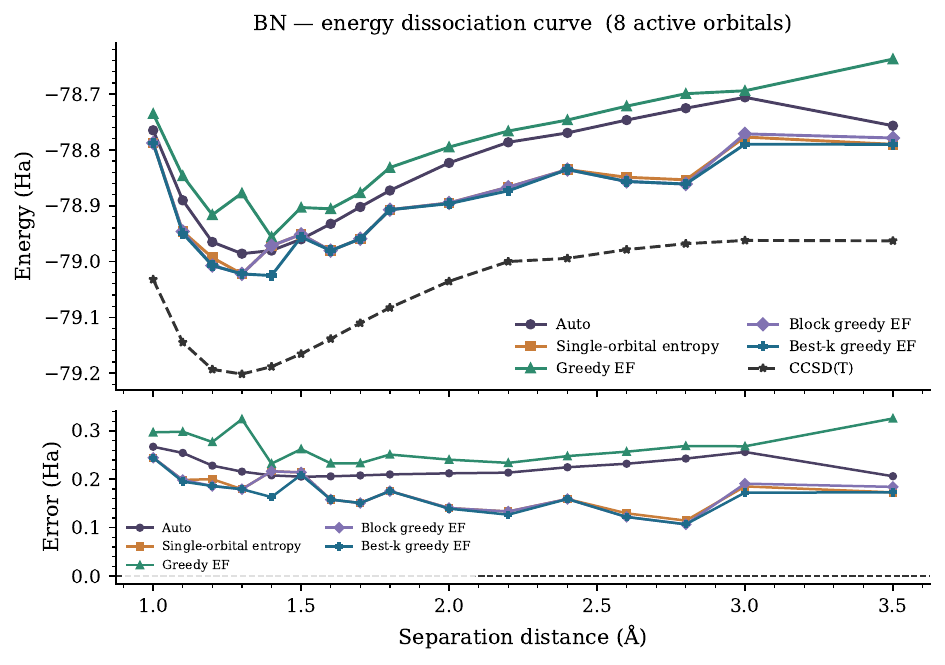}
        \caption{}
        \label{fig:energy_curve2}
    \end{subfigure}

    \caption{Energy dissociation curves for (a) $BeO/6-31G*$, and (b) $BN/6-31G*$. We plot automatically selected orbitals (navy circles), single-orbital entropy ranked orbitals (orange squares), greedy EF selected orbitals (green triangles), block greedy EF selected orbitals (purple diamonds), and best-$k$ greedy EF selected orbitals (blue pluses). The dotted line is a reference CCSD(T) calculation.}
    \label{fig:energy_curves}
\end{figure*}

The previous section shows the TNASS methods performing well in regions of high static correlation. We now demonstrate the performance of the methods along the entire dissociation curve. For the same $BeO/6-31G^*$ system, as well as $BN/6-31G*$, we plot the CASCI energy for each method at $n=8$ active orbitals compared to a CCSD(T) reference calculation. These are shown in Figure~\ref{fig:energy_curves}. We observe significantly better performance from the quantum information informed active spaces at almost all points of the dissociation curve. While the more advanced TNASS approaches yield almost identical energies to AutoCAS in the $BN$ system, we note the superior performance of the best-$k$ greedy methods in the $BeO$ system, particularly at stretched geometries. 

While total energies are often relatively insensitive to moderate deficiencies in the chosen active space due to their variational nature, molecular properties that depend explicitly on the electronic density, such as dipole moments, can exhibit much greater sensitivity to the quality of the active space \cite{kaufold2023automated}. Hence for $BN/6-31G*$, whose significant multi-reference character makes dipole moment calculations difficult, we plot the dipole moment calculated using the CASCI wavefunction obtained for each chosen active space compared to the dipole moment calculated from the reference CCSD(T) wavefunction. This is shown in Figure~\ref{fig:dipole_curve}.

\begin{Figure}
    \centering    
    \includegraphics[width=0.9\linewidth]{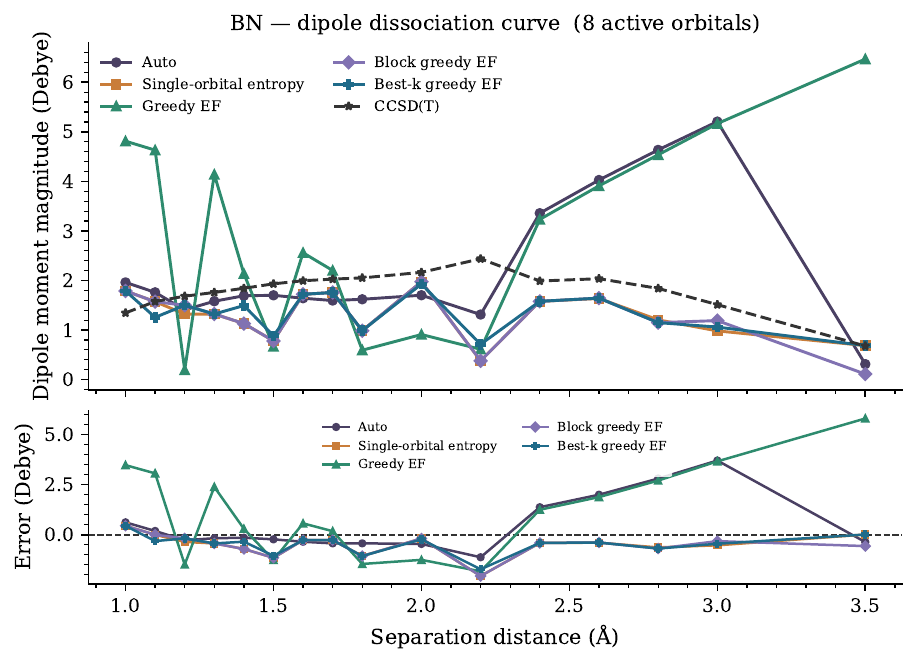}
    \captionof{figure}{Dipole moment dissociation curves for $BN/6-31G*$ calculated using the wavefunctions obtained in Figure~\ref{fig:energy_curve2}.}
    \label{fig:dipole_curve}
\end{Figure}

The figure shows that while orbitals chosen around the HOMO/LUMO gap provide a good wavefunction for dipole moment calculation at small separation distances, it fails significantly at larger separation distances ($2.4-3.0\AA$). The quantum information informed orbitals (with the exception of Greedy EF), remain stable throughout the dissociation.

\section{Conclusion}

We have introduced TNASS, a family of fully automated active space selection methods built on the Entanglement Feature, which uses multi-orbital Rényi-2 entropy evaluated via tensor network contraction to identify correlated orbital subspaces without requiring chemical intuition or target atom preselection. By framing single-orbital entropy selection (AutoCAS) as the $k=1$ limit of the block greedy family, we showed that the TNASS methods form a continuum of increasing computational cost and increasing sensitivity to multi-orbital entanglement structure, with the cost of each additional level of sophistication quantified analytically and validated empirically. 

Across the $BeO/6-31G*$ and $BN/6-31G*$ dissociation curves, the entanglement feature methods, particularly block greedy and best-$k$ greedy, matched or outperformed both the default HOMO/LUMO selection and single-orbital entropy ranking at compressed and stretched geometries, where correlation is diffuse across many orbitals, while remaining competitive at intermediate geometries where the HOMO/LUMO gap dominates. Best-$k$ greedy was the most consistently strong performer of the methods tested, though this comes at the highest computational cost among the variants considered. Plain greedy selection was consistently the weakest of the entanglement feature methods, a result which suggests that the lack of a submodularity guarantee for multi-orbital Rényi-2 entropy allows greedy to become trapped in local optima that later orbital choices cannot correct, and which motivated our development of block greedy and best-$k$ as more robust alternatives. Block greedy provides the best trade-off between performance and computational cost. 

Several open questions remain. Our results are presently limited to relatively simple diatomic systems; validating TNASS across a broader range of molecules, including larger and more strongly correlated systems such as transition metal complexes, is a natural next step and is where we expect the advantages of multi-orbital entanglement measures over single-orbital entropy to become more pronounced, since larger systems offer more scope for correlations that are invisible to any individual orbital's marginal entropy. Relatedly, the direct construction of the EF used throughout this work scales as $\mathcal{O}(\chi^4)$ in the underlying MPS bond dimension, restricting us to small $\chi$; adopting the Tensor Cross Interpolation approach of Kolisnyk et al. \cite{ef} may allow TNASS to be applied to systems where a larger bond dimension is required to capture the relevant entanglement structure, and represents a promising direction for extending the method's applicability.

A further direction is the integration of TNASS with molecular orbital optimisation methods. As discussed in Section 2.1, active space selection and orbital optimisation address complementary weaknesses: selection methods including TNASS choose the best subset from a fixed reference orbital set, while optimisation methods such as CASSCF improve the orbitals themselves but typically require a reasonable starting active space to converge reliably. A natural extension of this work would use TNASS-selected active spaces as the initial reference orbitals for AEGISS \cite{AEGISS}, replacing its single-orbital entropy preselection step with the multi-orbital entanglement information already computed by the EF, before applying AEGISS's atomic orbital projection and post-processing optimisation. This would combine the improved active space quality demonstrated here with the benefit of subsequent orbital refinement, and would be a direct test of whether TNASS's gains over single-orbital entropy selection persist, and compound, once orbital optimisation is introduced. More broadly, we intend to explore the automated cutoff parameter $\epsilon$ described in Section 3.3, which would allow TNASS to determine the active space size automatically rather than requiring $n$ as an input, further reducing the technical barrier to performing multireference calculations.

\section*{Conflicts of Interest}
The authors declare no conflicts of interest.

\section*{Data Availability}
The TNASS methods presented in this work are implemented in the open-source package \texttt{TN4QA} (available through PyPI or on GitHub: \url{https://github.com/UCL-CCS/TN4QA}). The notebook used to generate the results in this paper is available upon request to the corresponding author.

\section*{Acknowledgments}
 A.M.\ acknowledges funding from the Engineering and Physical Sciences Research Council (grant number EP/S021582/1). I.H. thanks EPSRC, University College London, and A*STAR for the funding of her PhD studentship. P.V.C.\ is grateful for funding from the European Commission for VECMA (800925) and EPSRC for SEAVEA (EP/W007711/1).

\printbibliography

\end{multicols}

\end{document}